\documentclass{ws-p8-50x6-00}

\newcommand{\JPR}[4]{{\it Phys. Rev.} #1 {\bf #2}, #3 (#4)}
\newcommand{\JNPB}[3]{{\it Nucl. Phys.} B {\bf #1}, #2 (#3)}
\newcommand{\JAP}[3]{{\it  Ann. Phys.} {\bf #1}, #2 (#3)}

\begin{document}

\title{The Giessen Model - Vector Meson Production on the Nucleon in a
        Coupled Channel Approach}

\author{G. Penner$^*$ and U. Mosel}

\address{Institut f\"ur Theoretische Physik, Universit\"at Giessen, D-35392 
Giessen \\$^*$Email: gregor.penner@theo.physik.uni-giessen.de}

\maketitle

\abstracts{ In ref.\cite{feusti} we developed a unitary gauge
invariant effective Lagrangian model including the final states $\gamma
N$, $\pi N$, $2\pi N$, $\eta N$, $K \Lambda$, and $K \Sigma$
(ref.\cite{agung}) for a simultaneous analysis of all avaible experimental
data for photon- and pion-induced reactions on the nucleon. In 
ref.\cite{keil} this analysis was extended to $K^-$ induced reactions. 
In this paper we discuss an extension of this method to vector meson nucleon
final states, outline a generalization of the standard partial wave formalism, 
that is applicable to any meson-/photon-baryon reaction, 
and show first results for $\omega N$ production.}

\section{Introduction}

The determination of nucleon resonance properties from experiments 
where the nucleon is excited either via hadronic or electromagnetic 
probes is one of the major issues of hadron physics. The goal is 
to be finally able to compare the extracted masses and partial decay widths 
with predictions from lattice QCD (e.g. ref.\cite{flee}) and/or quark models 
(e.g. ref.\cite{capstick}).

\begin{figure}[t]
  \begin{center}
    \epsfxsize=13pc
    \parbox{55mm}{\epsfbox{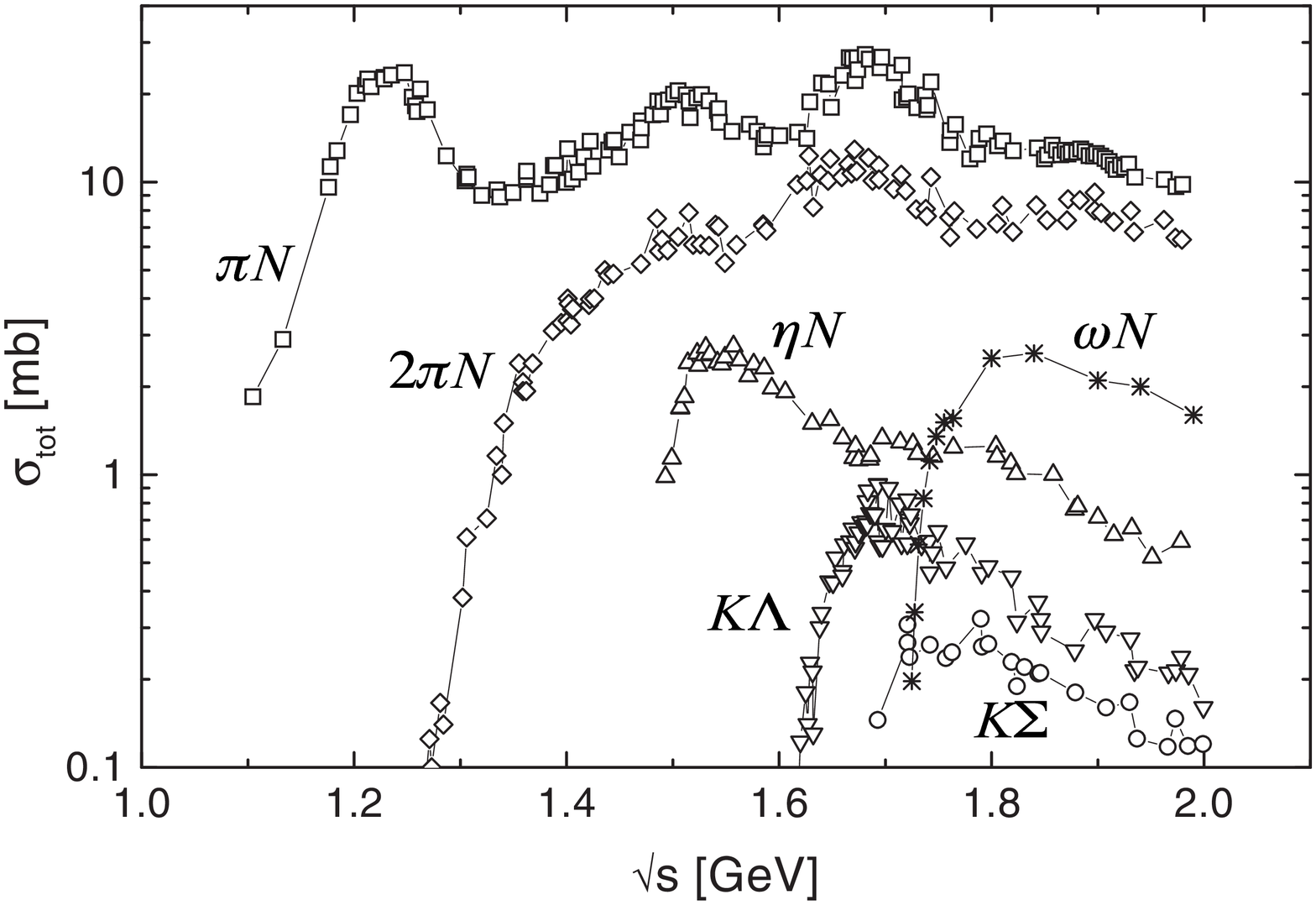}}
    \epsfxsize=13pc
    \parbox{55mm}{\vspace*{-2mm}\epsfbox{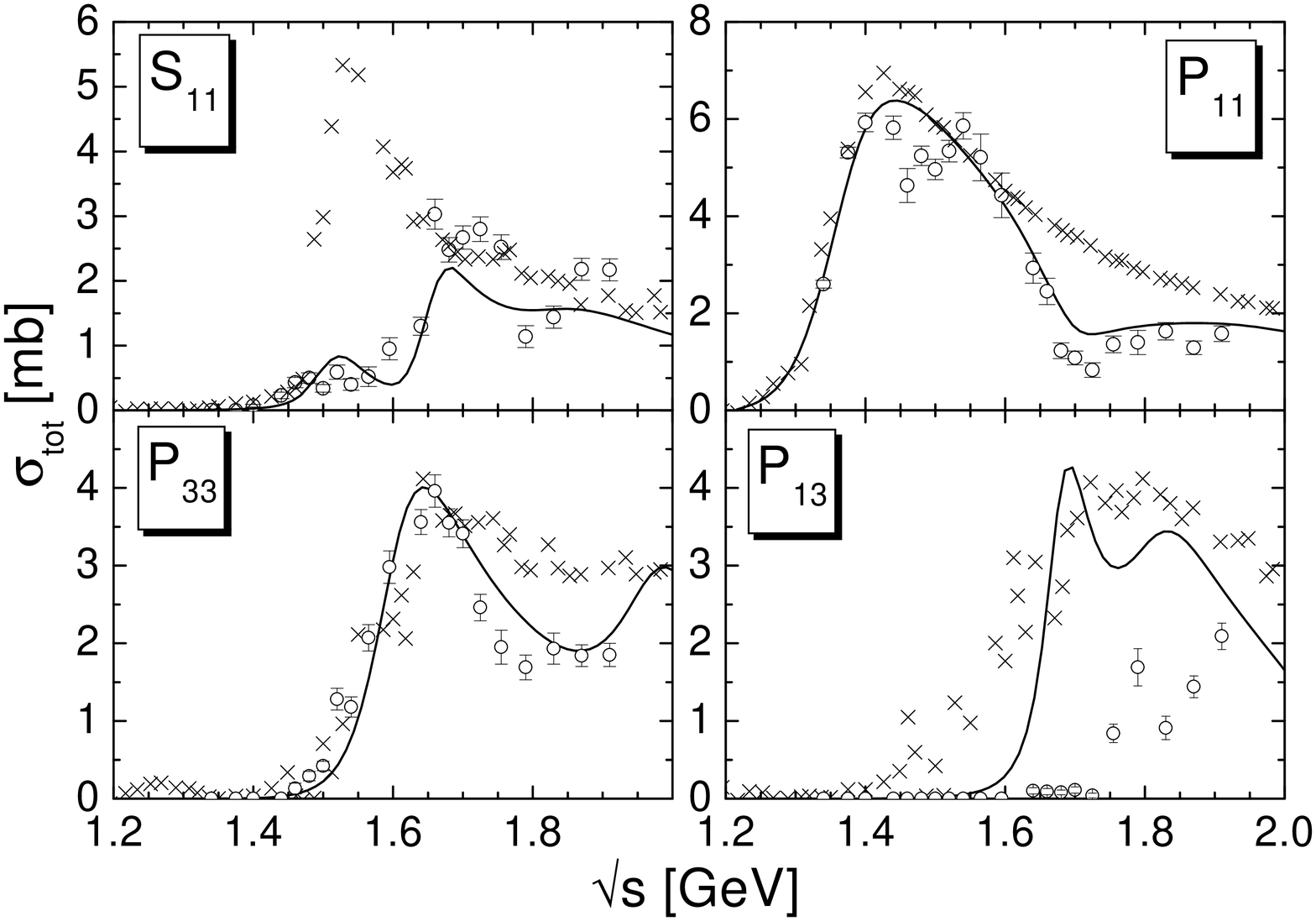}}
    \caption{Left: Total cross sections for the reactions $\pi^-p \rightarrow X$ 
      with $X$ as given in the figure. Data are from ref. $^6$. Right: 
      Total partial wave cross sections for $\pi N \rightarrow 2 \pi N$.
      \label{totand2pi}}
  \end{center}
\end{figure}

As has been shown in ref.\cite{feusti} for a reliable extraction of these 
properties it is inevitable to analyze photon and pion induced experimental 
data  simultaneously for as many channels as possible. Therefore our 
coupled channel model developed in ref.\cite{feusti} incorporates 
the final states $\gamma N$, $\pi N$, $2\pi N$, $\eta N$, and $K \Lambda$, 
where the $2\pi N$ channel was modelled for simplicity by an isovector 
$0^+$ meson. But as soon as we try to extend the model to CMS energies up to 
$\sqrt s = 2$ GeV for an investigation of higher and 
so-called missing nucleon resonances, the inclusion of the $\omega N$ final state 
becomes unavoidable due to unitarity. This can be seen from the left panel of 
Fig. \ref{totand2pi}.
Furthermore, $\omega$ production on the nucleon represents 
a possibility to project out $I = \frac{1}{2}$ resonances in the reaction 
mechanism. Due to its intrinsic spin the inclusion of the $\omega N$ final state 
in our coupled channel model requires an extension of the standard partial wave 
decomposition (PWD) method developped for $\pi N/\gamma N \rightarrow \pi N$ and 
$\gamma N \rightarrow \gamma N$ (see e.g. ref.\cite{feusti}). Such an extension
is provided in section \ref{vector}. In addition, this formalism enables us to 
achieve a realistic treatment of the most dominant inelastic channel in $\pi N$ 
scattering, i.e. the $2\pi N$ state via $\rho N$, $\pi \Delta$, and $\sigma N$.

\section{The Model}

Our method to solve the Bethe-Salpeter (BS) equation is the so called $K$ matrix 
Born-approximation, which is equivalent to setting the intermediate particles 
in the BS propagator on-shell; for more details cf. ref.\cite{feusti}. Then the 
reaction matrix $T$, defined by $S = 1 + 2\mbox i T$, can be calculated from the 
potential $V$ after a PWD in total spin $J$, parity $P$, and isospin $I$ via 
matrix inversion:
\begin{equation}
T(p',p;\sqrt s) = \frac{V(p',p;\sqrt s)}{1 - \mbox i V(p',p;\sqrt s)} .
\end{equation}
and unitarity is fulfilled as long as $V$ is hermitian.

The potential $V_{fi}$ is built up by $s$-, $u$-, and $t$-channel 
Feynman diagrams by means of effective Lagrangians which can be found in 
refs.\cite{feusti,keil}. The advantage of this method is that the background 
contributions are created dynamically and the number 
of parameters is greatly reduced, i.e. as compared to Breit-Wigner 
driven models or those including only pointlike interactions.

\section{Results on (pseudo)scalar meson production}

As an example for the quality of the calculations we show in the right 
panel of Fig. \ref{totand2pi} the total partial wave cross sections, as 
extracted by the standard PWD, for $\pi N \rightarrow 2 \pi N$ in comparison 
with the inelasticities from $\pi N \rightarrow \pi N$ ($\times$) and a 
$\pi N \rightarrow 2 \pi N$ analysis\cite{manley} ($\circ$). The necessity 
of the inclusion of a large set of final states in a coupled channel 
calculation can be seen in various partial waves. 
In the $S_{11}$ partial wave the difference between the inelasticity and the 
$2\pi$ analysis around $\sqrt s = 1.5$ GeV is easily explained by the 
opening of the $\eta N$ final state, the same holds true for $P_{11}$ 
above $\sqrt s = 1.6$ GeV and $K \Lambda$ and for $P_{33}$ 
above $\sqrt s = 1.7$ GeV and $K \Sigma$. However, there still 
is a discrepancy left in the $P_{13}$ partial wave arising around $1.7$ GeV. 
Since this particular calculation did not include the $\omega N$ final state 
yet, this problem might be solved upon a reanalysis including the $\omega N$ 
final state.

\section{Vector Meson Production} \label{vector}

Since the orbital angular momentum $\ell$ is not conserved in, e.g., 
$\pi N \rightarrow \omega N$ the standard PWD becomes really clumsy for 
many of the channels that have to be included. A more elegant and in 
particular uniform PWD for all channels would be desirable. Hence we use 
here a generalisation of the standard PWD method which represents a tool to 
analyze any meson- and photon-baryon reaction on an equal footing. 

We start with the decomposition of a two-particle momentum states into 
states characterized by the total spin $J$ and its $z$-component $M$ 
(see ref.\cite{jacobwick}):
\begin{eqnarray}
|p;JM,\lambda_1 \lambda_2 \rangle = 
  \sqrt \frac{2 J + 1}{4 \pi} \int e^{i(M-\lambda)\varphi} 
  d^J_{M \lambda}(\vartheta) 
  |p \vartheta \varphi;\lambda_1 \lambda_2 \rangle \mbox d \Omega, 
  \; \; \; \lambda = \lambda_1 - \lambda_2, \nonumber
\end{eqnarray}
where $\lambda_1$ and $\lambda_2$ are the helicities of the two particles and
the $d^J_{M \lambda}(\vartheta)$ are Wigner functions. For the incoming CMS 
state ($\vartheta_0 = \varphi_0 = 0$ $\Rightarrow$ $\ell_z = 0$) one gets 
$\langle JM,\lambda_1 \lambda_2 | \vartheta_0 \varphi_0 ,\lambda_1 
\lambda_2 \rangle \sim \delta_{M \lambda}$, hence $M = \lambda$ and one can 
drop the index $M$. By using the parity property (cf. ref.\cite{jacobwick}) 
$\hat P |J,\lambda \rangle = \eta_1 \eta_2 
(-1)^{J-s_1-s_2} |J,-\lambda \rangle$, where $\eta_1$, $\eta_2$, and 
$s_1$, $s_2$ are the intrinsic parities and spins of the two particles, 
the construction of states with a well defined parity is straightforward:
\begin{eqnarray}
  \hat P |J,\lambda;P\pm \rangle := 
    \hat P \frac{1}{2} (|J,+\lambda\rangle \pm |J,-\lambda \rangle) =
    \pm \eta_1 \eta_2 (-1)^{J-s_1-s_2} |J,\lambda; P\pm\rangle, \nonumber
\end{eqnarray}
and we can use them to project out helicity amplitudes with definite parity:
\begin{eqnarray}
  T^{J\pm}_{\lambda' \lambda} :&=& 
  \langle \lambda' | \hat T | J\lambda;P\pm \rangle \\
  \hspace*{3mm}&=& \frac{1}{2}  
    (T^J_{\lambda' \lambda} \pm T^J_{\lambda' -\lambda})
    \hspace{5mm} 
    \mbox{with } 
      T^J_{\lambda' \lambda} = 
      \frac{1}{2} \int T_{\lambda' \lambda}(x) 
      d^J_{\lambda \lambda'} (x) \mbox d x, \; x = \cos \vartheta . \nonumber
\end{eqnarray}
These helicity amplitudes $T^{J\pm}_{\lambda' \lambda}$ have definite, identical 
$J$ and definite, but opposite $P$. As is quite obvious this method is valid 
for any meson-baryon final state combination, even cases as e.g. 
$\omega N \rightarrow \pi \Delta$. In the case of $\pi N \rightarrow \pi N$ 
the $T^{J\pm}_{\lambda' \lambda}$ coincide with the conventional partial wave 
amplitudes: $T^{J\pm}_{\frac{1}{2} \frac{1}{2}} \equiv T_{\ell\pm}$.

\begin{figure}[t]
  \begin{center}
    \epsfxsize=14pc
    \parbox{55mm}{\epsfbox{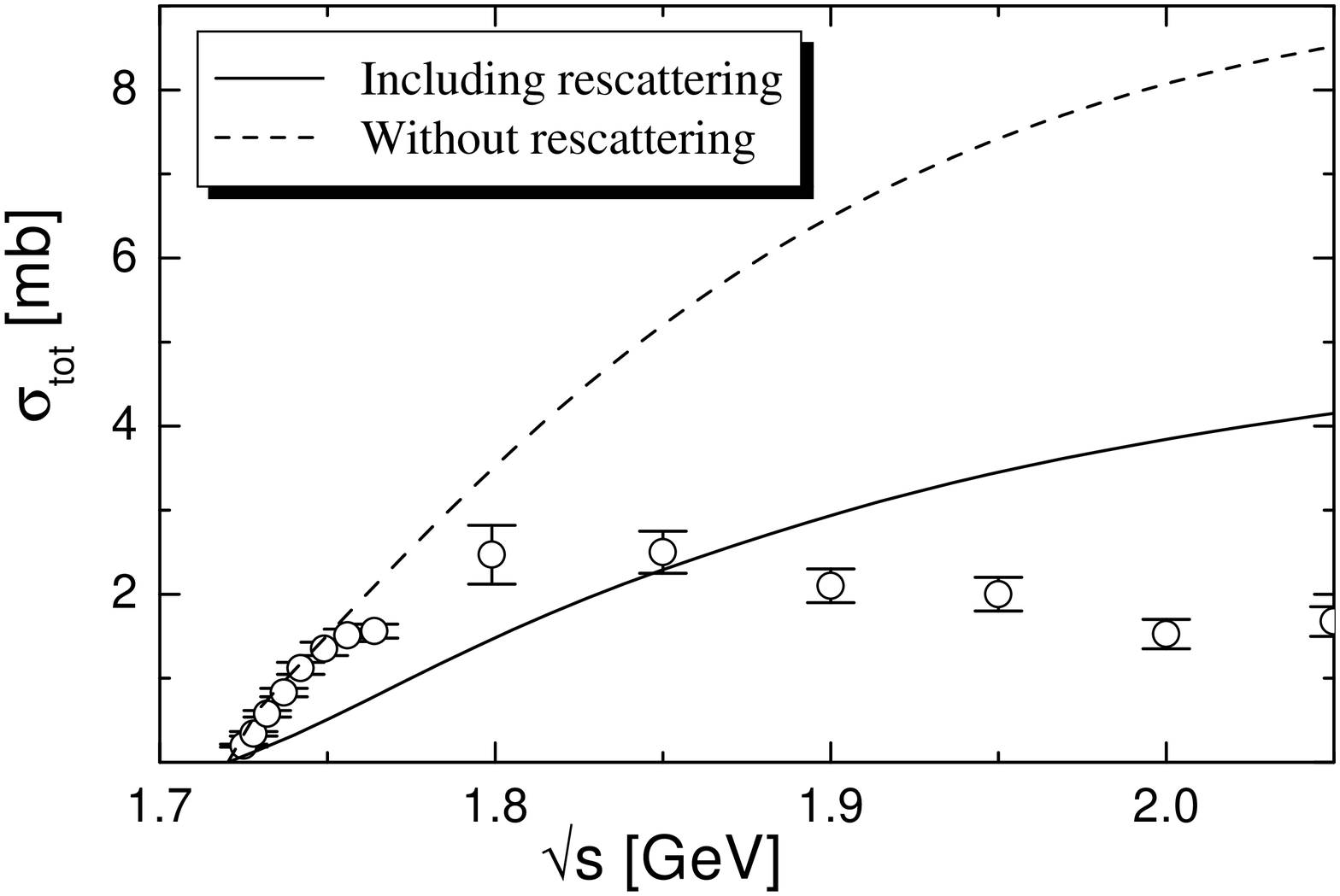}}
    \parbox{55mm}{\epsfbox{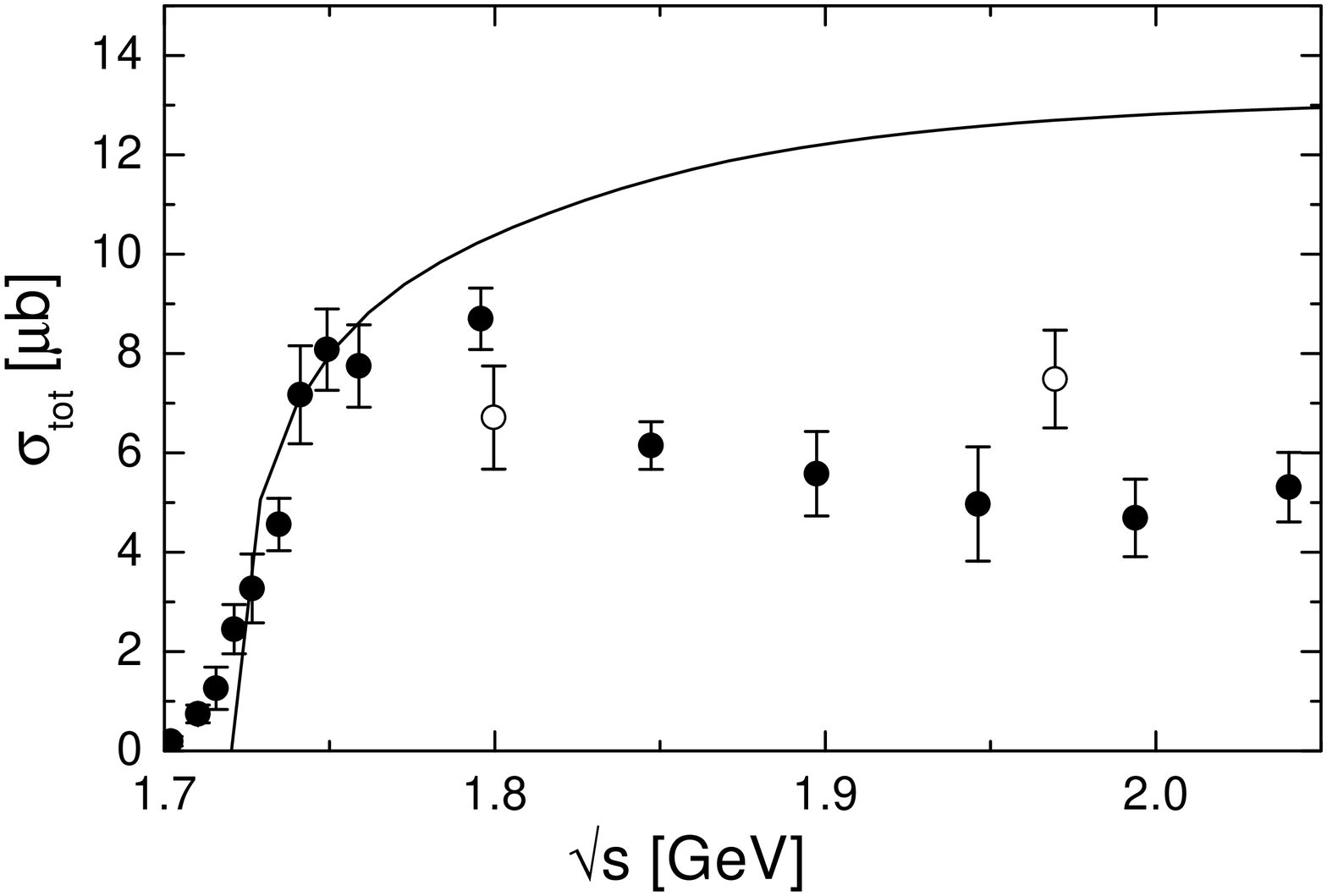}}
    \caption{First results for $\pi^- p \rightarrow \omega n$ (left) and 
      $\gamma p \rightarrow \omega p$ (right).\label{omega}}
  \end{center}
\end{figure}
This PWD has been used for calculating pion- and photon-induced $\omega$ 
production on the nucleon. For our first results, we have applied the couplings 
set SM95-pt-3 from ref.\cite{feusti}, i.e. the $\omega$ only couples to the 
nucleon. However, as can be seen in the left panel of Fig. \ref{omega} for a 
reliable calculation of $\omega$ production on the nucleon the inclusion of 
rescattering is a basic requirement.

\section{Outlook}

The next step of the extension of our coupled channel $K$ matrix model 
will be the inclusion of $\omega N$ data in the determination of 
resonance properties. Furthermore, since the partial wave formalism is now 
settled, the inclusion of additional final states, in particular for a more 
sophisticated description of the $2\pi N$ final state, such as $\rho N$, 
$\sigma N$, and $\pi \Delta$ and accounting for their spectral function 
is straightforward.

\section{Acknowledgments}

This work is supported by DFG and BMBF.

\end{document}